# Metal-free heteroatom doping of carbon nitride for enhanced photocatalytic hydrogen peroxide production

Eneko Sebastian-Garate[1], Bruna F. Gonçalves[2], Jordi Llusar[2], Pawel Gluchowski[3,4], Ivan Infante[2,5], Senentxu-Lanceros Mendez[2,5,6], Qi Zhang[2,5], Hugo Salazar[2,*]

[1]Departamento de Ingeniería Química, Facultad de Ciencia y Tecnología, University of the Basque Country (UPV/EHU), Sarriena s/n, 48940 Leioa, Spain

[2]BCMaterials, Basque Center for Materials, Applications and Nanostructures, UPV/EHU Science Park, 48940 Leioa, Spain

[3]Institute of Low Temperature and Structural Research PAS, PL-50422 Wrocław, Poland

[4]Graphene Energy Ltd., PL-50422 Wrocław, Poland

[5]IKERBASQUE, Basque Foundation for Science, 48013 Bilbao, Spain

[6]Physics Centre of Minho and Porto Universities (CF-UM-UP) and Laboratory of Physics for Materials and Emergent Technologies (LapMET), University of Minho 4710-057 Braga, Portugal

*Corresponding author: hugo.salazar@bcmaterials.net and lanceros@fisica.uminho.pt

## Abstract

The photocatalytic production of hydrogen peroxide ($H_2O_2$) from water is a promising approach for achieving solar-to-chemical energy conversion as a suitable alternative to obtain hydrogen fuel. Herein, we investigate the effect of heteroatom doping (B, O, P, and S) on the structural, electronic, and photocatalytic properties of graphitic carbon nitride ($C_3N_4$) for $H_2O_2$ production. A few-layered structure composed of stacked nanosheets was observed for pristine $C_3N_4$, while doping induced morphological and structural modifications, including increased disorder, exfoliation, and variations in interlayer spacing, indicating a successful heteroatom incorporation and perturbation of the electronic and optical properties. Photocatalytic tests under simulated sunlight demonstrate the influence of sacrificial agent, pH of the media, and the presence of scavengers on the $H_2O_2$ yield. A significant enhancement in $H_2O_2$ yield was achieved for all doped-catalysts, with 9.6, 14.8, 11.0, and 16.4-fold enhancement for B-, P-, O-, and S-doped $C_3N_4$, respectively, compared to the pristine catalyst. S-doped $C_3N_4$ catalyst achieved the highest yield of 3022.1 μmol $h^{-1}$ $g^{-1}$ and an apparent quantum yield (AQY) of 8.1%, due to improved charge separation and optimized selectivity. Mechanistic studies reveal that oxygen activation follows a $2e^-$ pathway, promoted by charge carrier modulation and reactive oxygen species (ROS) dynamics, with singlet oxygen and holes playing a dominant role. The stability of the catalysts is also demonstrated, with S-doped $C_3N_4$ maintaining over 95% efficiency after five cycles. These findings provide new insights into defect engineering for metal-free doping, offering a promising approach for photocatalytic, sustainable $H_2O_2$ generation.

**Keywords:** carbon nitride, defect engineering, heteroatom, metal-free doping, hydrogen peroxide, photocatalysis

## 1. INTRODUCTION

The global energy transition is a pressing challenge in the 21st century, driven by the need to reduce carbon emissions, mitigate climate change, and develop sustainable alternatives to fossil fuels. As industries and governments shift toward cleaner energy solutions, hydrogen-based technologies have garnered significant attention due to their potential for providing a versatile and renewable energy carrier [1]. However, the production, storage, and transport of molecular hydrogen present substantial technical and economic challenges, necessitating the exploration of alternative oxidative energy carriers [2]. Among these, hydrogen peroxide ($H_2O_2$) has emerged as a promising candidate, offering a safe, efficient, and scalable option for various applications.

Hydrogen peroxide is one of the most valuable chemicals in industry, and it is widely used in water treatment, chemical synthesis, medical disinfection, and environmental remediation industries [3]. It is also an ideal alternative to hydrogen ($H_2$) as chemical fuel, offering high energy density, zero emissions, and easy and safe storage and transport compared to compressed $H_2$ gas [4]. The global market size of $H_2O_2$ reached USD 3.07 billion by 2023, and it is expected to keep expanding at an annual growth rate of 5.7% until 2027 [5]. Currently, large-scale production of $H_2O_2$ relies on the well-established anthraquinone oxidation process, which is energy-intensive and environmentally harmful, generating exhaust gases and solid waste, and requiring expensive precious metal catalysts and organic solvents [6]. Thus, an environmentally friendly and cost-effective strategy to produce $H_2O_2$ is urgently needed to meet the growing demand for more sustainable $H_2O_2$ and align with global efforts to reduce carbon emissions.

Among the emerging methods for $H_2O_2$ synthesis, photocatalysis offers a highly attractive approach due to its ability to drive chemical reactions using sunlight. Photocatalytic $H_2O_2$ production involves the activation of catalysts under light irradiation, leading to the generation of electron-hole ($e^-/h^+$) pairs [7]. These charge carriers produce reactive oxygen species (ROS), which in turn promote both water oxidation reaction (WOR) and oxygen reduction reaction (ORR), resulting in $H_2O_2$ formation [8]. This process eliminates the need for hazardous chemicals and external energy input, allows the direct use of solar energy, requires mild reaction conditions, and has the potential scalability for decentralized applications, which makes it an eco-friendly and sustainable alternative to the current used anthraquinone process, which requires a high energy input, large infrastructures, and generates substantial waste [9]. Various semiconductor catalysts, such as titanium dioxide, metal oxides, and metal-free materials, such as graphene-based and porous carbon materials, have been explored for this purpose. However, developing efficient photocatalysts with high selectivity, stability, and activity remains a critical challenge in this field [10].

Among the various photocatalysts investigated for $H_2O_2$ production, carbon nitride ($C_3N_4$) has emerged as one of the most promising materials due to its metal-free composition, visible-light activity, charge carrier generation, and excellent chemical stability [11,12]. $C_3N_4$ exhibits a suitable conduction band potential for driving the $2e^-$ ORR, while surface defects and active sites play a key role in stabilizing reaction intermediates and promoting the subsequent formation of $H_2O_2$ [13]. Furthermore, its tunable electronic properties and ease of synthesis make it an attractive platform for designing advanced photocatalytic systems [14]. However, pristine $C_3N_4$ photoactivity suffers from limitations such as rapid charge recombination, suboptimal light absorption, and limited active sites, demanding further optimization strategies to enhance its photocatalytic performance [15]. To address these challenges, doping strategies with metal-free elements have been widely explored to enhance the electronic and structural properties of carbon nitride. The incorporation of heteroatoms such as phosphorus (P), oxygen (O), sulfur (S), and boron (B) can effectively enhance the $O_2$ adsorption, increase the charge carrier lifetime, and inhibit $H_2O_2$ decomposition by modifying active sites and charge polarization [16]. More specifically: P-doping alters the electronic configuration and shifts the $H_2O_2$ generation pathway from a $1e^-$ to a thermodynamic more favorable $2e^-$ pathway [17]; O-doping creates mid-gap states that facilitate charge transfer and enhance light absorption by $n \rightarrow \pi^*$ electron transition [18]; S-doping improves light absorption by narrowing the bandgap [19]; and B-doping results in an appropriate band structure, optimized charge transfer, and selective $2e^-$ ORR [14].

Due to the remarkable tunability and multifunctionality of $C_3N_4$ through heteroatom doping, this study aims to systematically investigate how such modifications influence photocatalytic performance. Specifically, the role of P, O, S, and B dopants is explored to enhance photocatalytic $H_2O_2$ production under sunlight. By advancing our understanding of $C_3N_4$-based systems, this research, complemented by Density Functional Theory (DFT) calculations, allows to unravel the underlying electronic structure and structural effects of doping, and contributes to the broader goal of developing sustainable photocatalytic platforms for green chemical production.

## 2. RESULTS AND DISCUSSION

### 2.1. Catalysts characterization

The morphology of $C_3N_4$ and its heteroatom-doped $C_3N_4$-based catalysts was analyzed through scanning electron microscopy (SEM) and transmission electron microscopy (TEM), revealing the distinct morphological modifications induced by B, O, P, and S doping (Figure 1a–b). Pristine $C_3N_4$ exhibited an aggregated, layered, graphitic-like structure with a relatively smooth surface, characteristic of its two-dimensional structure formed by tri-s-triazine units (Figure 1a1). The TEM image further confirmed the presence of a few-layered structure with stacked nanosheets, indicative of strong interlayer interactions (Figure 1b1). Upon heteroatom doping, minor morphological changes were observed. For B (Figure 1a2 and 1b2), P (Figure 1a4 and 1b4), and S (Figure 1a5 and 1b5) doping, the SEM and TEM images are similar to pristine $C_3N_4$. The layered, graphitic-like structure is kept and the degree of exfoliation remain in a few-layered magnitude. For O doping (Figure 1a3 and 1b5), TEM image suggests a more exfoliated morphology, showing thinner nanosheets with improved dispersion. Overall, the catalysts present similar morphologies, suggesting that the differences in $H_2O_2$ production cannot be ascribed to morphological factors. Differences in performance are related to the electronic modifications and defects introduced by heteroatom doping, which modulate charge separation and ROS dynamics.

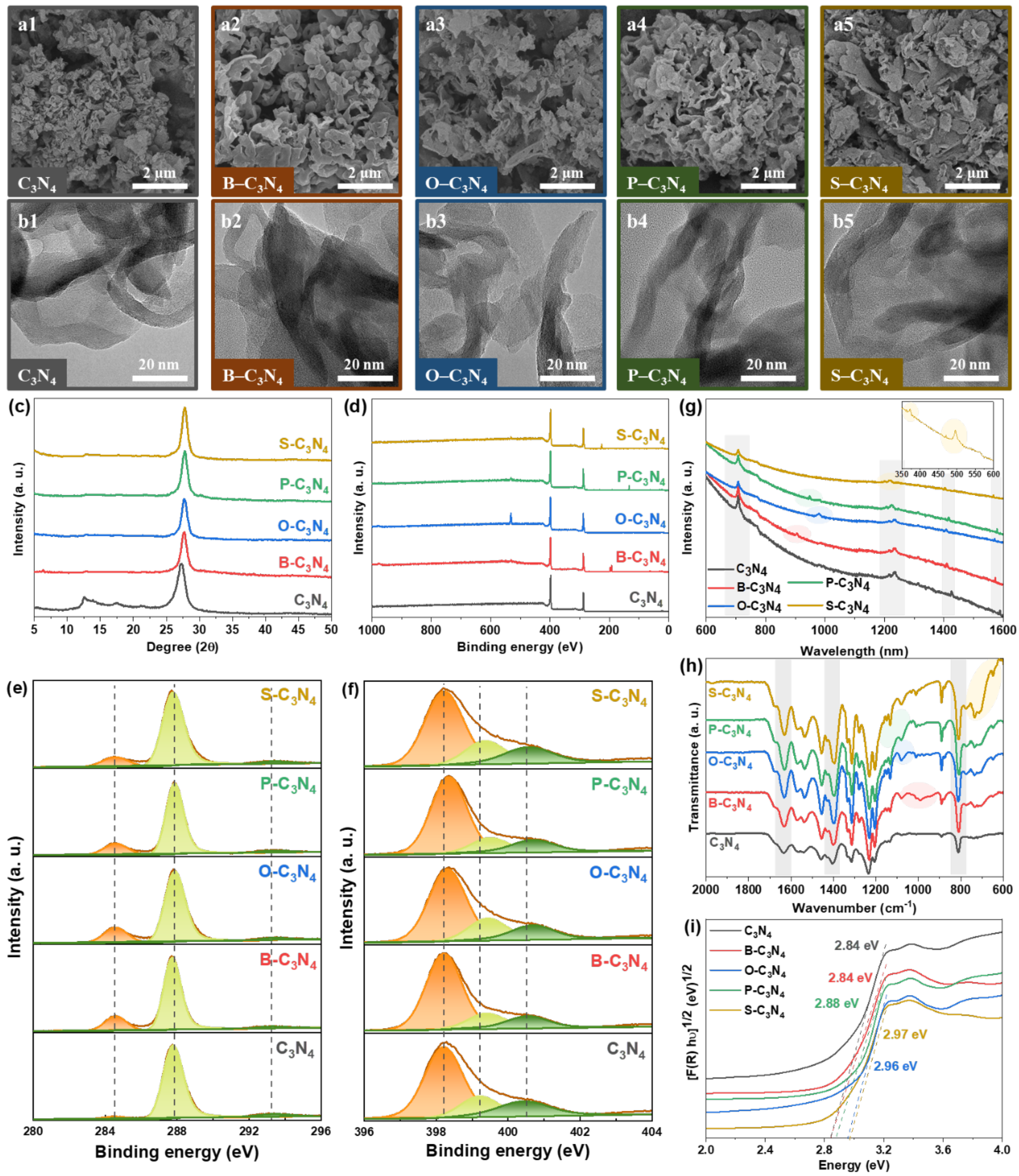

**Figure 1.** **(a)** SEM and **(b)** TEM images of **(a1, b1)** $C_3N_4$; **(a2, b2)** B-$C_3N_4$; **(a3, b3)** O-$C_3N_4$; **(a4, b4)** P-$C_3N_4$; **(a5, b5)** S-$C_3N_4$; **(c)** XRD patterns; **(d)** XPS full spectra; **(e)** C 1s spectra; **(f)** N 1s spectra; **(g)** Raman spectra; **(h)** FTIR spectra; **(f)** transformed Kubelka-Munk vs. hν plot of $C_3N_4$, B-$C_3N_4$, O-$C_3N_4$, P-$C_3N_4$, and S-$C_3N_4$.

X-ray diffraction (XRD) patterns of pristine $C_3N_4$ exhibit characteristic peaks at approximately 12.6º and 27.2º, corresponding to the in-plane structural packing of tri-s-triazine units (210) and interlayer stacking of the conjugated aromatic system (100), respectively (Figure 1c) [14]. The data refinement suggests the triclinic lattice planes with a space group *P1*. According to pattern matching refinements, estimated cell parameters of $C_3N_4$ and all doped catalysts are shown in Table S1. Upon doping with B, O, P, and S heteroatoms, the (100) peak at 27.2º shifts slightly to higher angles (Figure S1a), indicating an increase in interlayer spacing due to the larger atomic radius of these dopants compared to carbon and nitrogen, which is favorable for charge carrier mobility, increasing the $H_2O_2$ production [15]. The increase of the interlayer spacing is further confirmed by the pattern matching refinements (Figure S2). Apart from B-$C_3N_4$, all the other modifications enhanced the interlayer spacing (Table S1). The intensity of the (210) peak at 12.6° also decreases upon doping (Figure S1b), suggesting structural disorder and reduced crystallinity, particularly in boron and phosphorus-doped catalysts [20]. It is also reported that the replacement of C by these heteroatoms causes a reduction in the crystallinity, which implies a longer interlayer distance [21]. The heteroatom doping may disrupt the π-conjugated stacking of $C_3N_4$, introducing structural disorder and reducing the crystallinity. These disruptions affect the interlayer spacing, which in turn influences charge carrier mobility. A larger interlayer spacing promotes charge transfer, while enhanced disorder enhances $e^-/h^+$ recombination, impacting photocatalytic efficiency [22].

X-ray photoelectron spectroscopy (XPS) provides insights into the structure, elemental composition, and chemical states in $C_3N_4$ based catalysts (Figure 1d). Pristine $C_3N_4$ typically shows two main peaks at 287.3 and 398.1 eV, ascribed to C 1s and N 1s signals. It can also be noted the presence of a negligible peak around 530 eV, related to the O 1s signal. In the C 1s spectrum of $C_3N_4$ (Figure 1e), there are three component peaks located at 284.5, 287.7, and 293.3 eV, corresponding to C–C bonds, C–$NH_x$ groups, and N–C=N bonds, respectively [23]. The N 1s spectrum of $C_3N_4$ (Figure S3b) is fitted into three peaks located at 398.3, 399.2, and 400.5 eV, ascribed to bi-coordinated N(C–N=C), tri-coordinated N (N–$(C)_3$), and C–N–H bonds in the heptazine framework, respectively [24]. Doping induces changes, mainly in the increase in peak intensity ascribed to C–C bonds (284.5 eV) in the C 1s spectra (Figure 1e), while slightly shifting the peaks of N 1s (Figure 1f). Although the heteroatoms substitute C in the $C_3N_4$ structure, their introduction disrupts the heptazine framework, creating defects that induce a partial formation of C–C bonds or graphitic domains [25]. The shifts in the N 1s spectra are attributed to an electron density redistribution induced by the heteroatoms, which either withdraw electrons and modify the charge delocalization, leading to more electron-deficient N atoms [26]. For B-doped $C_3N_4$, one B 1s peak is observed at 191.7 eV, characteristic of $sp^2$-hybridized B and the formation of B–N bonds, a signature of substitutional doping of B into the framework by replacing N atoms indicating (Figure S4) [27]. For O-doped $C_3N_4$, the O 1s peak at 531.9 eV suggests that O heteroatoms have been doped into the $C_3N_4$ matrix (Figure S5) [28]. P-doped catalyst displays a P 2p peak near 133.0 eV, associated with P–N bonding (Figure S6) [29], while S-doped $C_3N_4$ shows a S 2p peak at 163.5 eV, which can be indicative of both C–S and N–S bonds, confirming the introduction of sulfur (Figure S7) [30].

The Raman spectrum of pristine $C_3N_4$ is mainly characterized by vibrational modes ascribed to its tri-s-triazine structure (Figure 2c). The breathing mode of triazine rings appears at 708 $cm^{-1}$, while skeletal C–N stretching modes are observed in the 1206–1255 $cm^{-1}$ range. Additionally, the observed D and G bands, located at 1426 and 1589 $cm^{-1}$, respectively, represent disorder and in-plane vibrations, similar to graphitic materials [31]. Doping with heteroatoms results in structural and electronic changes. B-doping introduces a new B–N stretching peak at 906 $cm^{-1}$ and causes a slight redshift of the peak at 706 $cm^{-1}$, due to changes in ring vibrations [21]. O-doping introduces new C–O and N–O vibrations (965–989 $cm^{-1}$) and increases disorder, broadening the D-band [31]. P-doping shifts skeletal C–N peaks to lower wavenumbers and introduces a P–N peak (948 $cm^{-1}$) [32]. S-doping weakens C–N interactions, leading to a downshift of C–N peaks and the appearance of C–S and N–S vibrations (372 and 496 $cm^{-1}$) [33]. Overall, doping increases disorder – as noted by the increase of the ratio between the D and G bands ($I_D/I_G$). Pristine $C_3N_4$ shows an $I_D/I_G$ ratio of 0.96. By doping with B, O, P, and S, these values increase to 1.19, 1.57, 1.09, and 1.58, respectively, indicative of structural disorder and defect formation.

The Fourier-transform infrared spectroscopy (FTIR) spectra of pristine and heteroatom-doped $C_3N_4$ provide valuable insights into the structural modifications induced by B, O, P, and S doping (Figure 1h). $C_3N_4$ exhibits characteristic absorption bands in the range of 1000–1800 $cm^{-1}$, attributed to the stretching vibrations of heptazine framework – at 809, 1403, and 1634 $cm^{-1}$, ascribed to s-triazine-s rings, C–N, and C=N stretching vibrations, respectively [34] –, along with a broad band around 3000–3500 $cm^{-1}$ due to N–H and O–H stretching from residual surface functional groups and adsorbed water (Figure S8). Upon doping, relevant changes can be observed. A red shift is observed across all doped materials for the C–N and C=N stretching bands due to electronic interactions between heteroatom and the $C_3N_4$ framework and charge redistribution within the framework. Also, the higher intensity of these two bands indicates a more exfoliated structure [35]. B-$C_3N_4$, has a new absorption region between 940 and 1010 $cm^{-1}$, corresponding to B–N and B–O stretching vibrations [36]. O-doping introduces perturbations that influence bond vibrations, enhancing the intensity of the broad O–H band due to increased surface hydroxyl groups [35]. A distinctive feature of O-$C_3N_4$ is the presence of additional C–O stretching vibrations around 1000–1100 $cm^{-1}$, suggesting partial oxidation of the structure [14]. P-doping induces new P–N and P=O stretching vibrations between 1040 and 1170 $cm^{-1}$, and influence hydrogen bonding interactions, modifying the intensity of N–H and O–H vibrations [29]. S-doping significantly increases intensity in the region of 600–900 $cm^{-1}$, associated with C–S or S–N bonds [37], and affects the hydrogen bonding network, increasing the intensity of the absorption bands located around 3000–3500 $cm^{-1}$.

The optical band gap of pristine $C_3N_4$, determined from UV-Vis diffuse reflectance spectroscopy, is 2.84 eV (Figure 2D), which agrees with previous works [34]. Doping with heteroatoms influences the band gap due to modifications in the electronic structure of the catalyst. B-doped $C_3N_4$ retains a band gap of 2.84 eV, indicating minimal disturbance to the conjugated system [38]. In contrast, O-doping increases the band gap to 2.97 eV due to the strong electronegativity of oxygen, which withdraws electron density and induces a blue shift, consistent with previous reports [34]. P-doped catalyst shows a band gap of 2.88 eV, slightly higher than pristine $C_3N_4$, attributed to the introduction of lone pair electrons from phosphorus, which modifies the conduction band [39]. S-doped $C_3N_4$ has a band gap of 2.96 eV, aligning with literature findings reporting that sulfur's larger atomic size and weaker electronegativity lead to structural distortion and electronic state changes, alongside the sulfur-induced extension of the conjugated π-systems and creation of impurity states near the conduction band [40,41]. These band gap variations highlight how heteroatom doping allows to tailor the electronic properties of $C_3N_4$, enhancing its suitability for photocatalytic applications.

Figure S9 presents the $N_2$ adsorption-desorption isotherms of $C_3N_4$, B-$C_3N_4$, O-$C_3N_4$, P-$C_3N_4$, and S-$C_3N_4$. In the case of $C_3N_4$, a faint hysteresis loop is observed, suggesting a poor pore structure. The doping of $C_3N_4$ introduced significant changes in the porous structure – the catalysts present a type IV isotherm with an H3 hysteresis loop, suggesting the presence of mesopores [42]. The BET specific surface areas of $C_3N_4$, B-$C_3N_4$, O-$C_3N_4$, P-$C_3N_4$, and S-$C_3N_4$ are calculated to be 61.6, 67.3, 102.1, 78.2, and 141.5 $m^2\ g^{-1}$, respectively (Table S3). The higher surface upon doping suggests a higher number of catalytic sites for the $2e^-$ ORR, ultimately leading to higher $H_2O_2$ yields.

## 2.2. Theoretical calculations

To gain deeper insights into the effects of doping $C_3N_4$ with B, P, O and S atoms, density functional theory (DFT) calculations were performed. The employed unit cell represents a super cell containing a complete layer and two additional half-layers on the top and the bottom, representing a total of two layers. The unit cell employed exhibits the commonly-observed staggered structural configuration among layers. The out-of-plane size is approximately 0.9 nm, while the in-plane dimensions are 1.3 nm and 2.3 nm (Figure 2a). Further details can be found in Section 2 of the Supporting Information.

The pristine $C_3N_4$ model features a clean HOMO-LUMO gap with no trap states (Figure 2b-i-top). This is corroborated by the delocalized wavefunction of the frontier molecular orbitals (Figure 2b-i-bottom), where conjugated bonds within the heptazine units facilitate electron and hole delocalization throughout the nanosheets. Additionally, the LUMO and the nearly degenerate LUMO+1 show π-bonding character between C and N *p* atomic orbitals all along the nanosheets featuring the double bonds present within the heptazines. Instead, HOMO prompts π-antibonding behavior between C and N *p* atomic orbitals. In structural terms, $C_3N_4$ slightly displays some bending, yet with well-defined crystallinity (Figure S10a), which helps to promote a good delocalization of electrons and holes throughout the entire system.

Introducing dopants can affect the properties of the pristine lattice, and depending on their atomic size, coordination, and electronegativity, charge carriers may experience localization near the dopant site. Additionally, structural reorganization is expected based on the specific dopant introduced at specific sites into the lattice. To assess how this may impact the electronic properties and stability of the pristine $C_3N_4$ model, we explored systematically various representative lattice sites for substitution (Figure 2c), simplifying the analysis by introducing a single dopant.

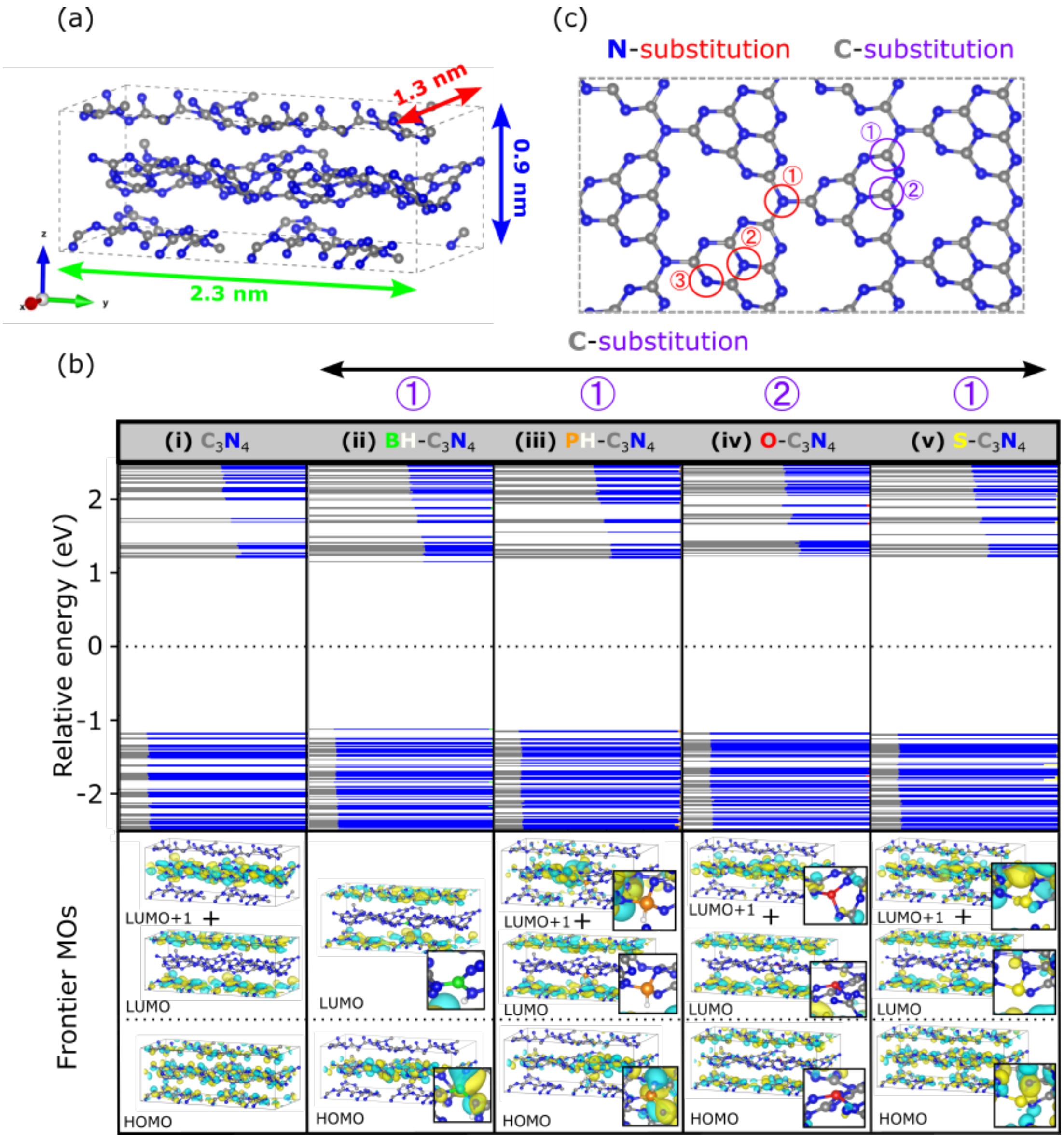

**Figure 2. (a)** Sketch of the $C_3N_4$ model used in the DFT calculations. **(b)** (Top panels) Projected density of states (PDOS) and (Bottom panels) frontier molecular orbitals (MOs) for ***(i)*** $C_3N_4$, ***(ii)*** B-$C_3N_4$, ***(iii)*** P-$C_3N_4$, ***(iv)*** O-$C_3N_4$ and ***(v)*** S-$C_3N_4$. As for the PDOS plots, the horizontal segments represent the MOs forming the electronic structure of the material and the colors within the segments show the contribution of each kind of element to these MOs; the sum of all contributions add up to 1 (or 100%). The horizontal black dotted line in top and bottom panels denotes the Fermi level. In the bottom panels are depicted zoomed-in pictures of the dopant. When both the LUMO and LUMO+1 are explicitly shown, it indicates that these molecular orbitals are nearly degenerate in energy and can be treated as a single effective orbital, denoted as LUMO′, which represents the electrons' ground state as a linear combination of the two; **(c)** Representative (left) N-substitution and (right) C-substitution sites all along the monolayers.

B atoms are usually expected to replace N atoms [43], while P, O and S are energetically favored for substituting C sites [34,37]. However, it is important to point out that, based on electronegativity, B is more similar to C than N, and may also replace this atom in the heptazine ring. For B-atom doping, thus, we also analyzed substituting C with B, however a simple swap of the atoms leads to a removal of one valence electron, which gives rise to an unstable B radical. To circumvent this, we introduced B in BH form, which is expected to be more energetically favorable and is isoelectronic with C. For the most stable substitution in ①, B binds to its neighboring N atoms, forming B-N bonds analogous to the C-N bonds at the ①-site in $C_3N_4$. The crystallinity of the layer remains largely preserved after B(H)-doping, although the monolayer exhibits a slightly flatter morphology (Figure S10b). Thus, the incorporation of B(H) into C sites would align with the B-N bonds detected by XPS and FTIR analyses (Figure S4 and Figure 1h). Inserting a single B(H) into a C-site does not introduce substantial modifications in terms of HOMO – LUMO gap and carrier delocalization, however, electrons and holes occupy separate layers (Figure 2b-ii). Contrary to this kind of substitution, when introducing B into the N-site,  the initially delocalized MOs become localized, concentrated around the B atom as observed in the PDOS plot (Figure S11b-bottom).

Upon P-doping, as P is expected to substitute a C, this introduces one extra valence electron compared to C, which would translate into a radical $P^{\bullet}$ species. However, this scenario is energetically unlikely. Thus, we expect P to enter the $C_3N_4$ lattice in PH form, ensuring a more stable closed-shell electronic configuration. Substituting P in the form of PH at site ① is the most stable configuration (Table 1) and does not appear to cause substantial structural and electronic modifications (Figure S12), with only a slight localization of the HOMO and LUMO (Figure 2b-iii). Therefore, even at low doping concentrations, one can expect localization features in such systems.

In the case of O doping, the oxygen atom substitutes a C atom, slightly favoring site ②, although substitution at site ① is close in energy and could also be accessible (Table 1). When O is incorporated, some in-plane deformations are induced (Figure S10d). Despite this structural deformation, HOMO and LUMO remains unaltered (Figure 2b-iv). Instead, when O is placed at site ①, the LUMO wavefunction becomes localized around the O atom, while the HOMO wavefunction retains its delocalized character (Figure S13).

Regarding S atoms substitution, they can replace either N or C in the $C_3N_4$ monolayers. In terms of stability, C-substitution has proved to be significantly favored over the N-substitution [37]. Within both representative C-substitution configurations, shown in Figure 2c**,** it exists a preference for replacing C by S at ① (Table 1). In terms of structural deformation, introducing S atoms seemingly distorts the $C_3N_4$ structural integrity by bending those nanosheets in-plane (Figure S10e). Furthermore, our calculations show that replacing C with S does not introduce any localized state, instead, they exhibit an even better delocalized LUMO' and HOMO wavefunctions than those observed for pristine $C_3N_4$ (Figure 2b-v). However, at high S doping levels, once the more favorable C-substitution sites are saturated, S atoms may start replacing N atoms (Figure S14). Although this type of substitution is less likely, it can lead to the emergence of localized HOMO and LUMO states, potentially introducing trap states.

**Table 1.** Relative energies corresponding to the configuration comparisons listed in the second column, for various atomic substitutions in $C_3N_4$. Each value reflects the energy difference between the most stable configuration and a less stable one. Substitutions include B (in BH form), P (in PH form), O, or S replacing C.

| Substitution | Configuration comparison | ΔE (kcal mol$^{-1}$) |
|---|---|---|
| B (BH) → C | ① BH-$C_3N_4$ – ② BH-$C_3N_4$ | 1.6 |
| P (PH) → C | ① PH-$C_3N_4$ – ② PH-$C_3N_4$ | 11.8 |
| O → C | ② O-$C_3N_4$ – ① O-$C_3N_4$ | 5.7 |
| S → C | ① S-$C_3N_4$ – ② S-$C_3N_4$ | 6.2 |

## 2.3. $H_2O_2$ generation experiments

To explore the effect of the developed materials on the photocatalytic generation of $H_2O_2$, the activity of $C_3N_4$, B-$C_3N_4$, O-$C_3N_4$, P-$C_3N_4$, and S-$C_3N_4$ was evaluated under simulated sunlight. As a first experiment, different types of sacrificial agents – water (no sacrificial agent), ethanol, benzyl alcohol, and isopropanol – were screened (Figure 3a). Taking as a basis the $H_2O_2$ generation using $C_3N_4$ as a catalyst in the presence of only water, a $H_2O_2$ yield of 30.7 µmol h$^{-1}$ g$^{-1}$ was obtained. By adding sacrificial agents to the solution, a significant increase in the production was observed: 73.6, 139.2, and 184.4 µmol h$^{-1}$ g$^{-1}$ for ethanol, benzyl alcohol, and isopropanol, respectively. The sacrificial agent plays a critical role in the reaction by donating electrons, acting as hole scavengers, and suppressing side reactions – ORR occurs via the 2e$^-$ pathway. Water alone provides limited electron donors, due to high recombination of charge carriers [35]. Ethanol acts as a hole scavenger by being oxidized to acetaldehyde. However, acetaldehyde can undergo further oxidation to acetic acid, introducing a 4e$^-$ side reaction (reduction of $O_2$ to $H_2O$ instead of $H_2O_2$) that hinders the selectivity of the 2e$^-$ WOR and results in $H_2O_2$ decomposition [44]. Benzyl alcohol as a hole scavenger enhances charge separation by being oxidized to benzaldehyde, which shows a higher oxidation potential than acetaldehyde and, therefore, being more difficult to undergo further oxidation. This results in a 4.5-fold enhancement in $H_2O_2$ yield; nevertheless, the lower kinetics of benzyl alcohol oxidation results in a moderate enhancement compared to isopropanol [45]. Isopropanol reached the highest $H_2O_2$ yield, inducing a 6-fold enhancement. This is attributed to its oxidation to acetone, a stable product that does not undergo further oxidation, preventing competing side reactions that could consume photogenerated charge carriers, and ensuring the availability of electrons for the ORR and WOR pathways. Moreover, isopropanol presents a lower oxidation potential than benzyl alcohol, which allows a faster and efficient hole scavenging process, further increasing the overall efficiency of $H_2O_2$ generation [46]. These results are consistent with the extent of oxidation products detected by carbonyl capture (DNPH) measurements (Figure S15): acetone > benzaldehyde > acetaldehyde. The higher concentration of acetone confirms the superior performance of isopropanol as sacrificial agent. In contrast, acetaldehyde is detected in the lowest concentration, which can be related to its conversion to acetic acid, which reduces $H_2O_2$ selectivity through $4e^-$ pathways [47].

The effect of the metal-free heteroatom doping on the $H_2O_2$ yield and kinetics was investigated using isopropanol as sacrificial agent, showing that this defect engineering strategy significantly influenced the photocatalytic $H_2O_2$ yield (Figure 3b). The pristine $C_3N_4$ exhibited the lowest performance (184.4 μmol $g^{-1}$ $h^{-1}$), highlighting its inherent limitations in charge separation and light absorption [15]. Among the doped catalysts, B-$C_3N_4$, O-$C_3N_4$, P-$C_3N_4$, and S-$C_3N_4$ exhibited significantly enhanced performances, with yields of 1771.3, 2730.1, 2031.0, and 3022.1 μmol $g^{-1}$ $h^{-1}$, respectively. The improved activity upon doping is attributed to the modulation of the electronic structure, enhanced charge separation, and increased active sites for ORR and WOR [9]. Boron doping plays a dual role by modulating band structure and introducing Lewis acid sites to enhance $O_2$ activation, enhancing charge separation, and favoring selectivity for the $2e^-$ ORR pathway, as reflected in the 9.6-fold increase in $H_2O_2$ yield. The mid-gap states created by O-doping promote charge transfer and light absorption through $n \rightarrow \pi^*$ electron transitions, resulting in a nearly 14.8-fold increase in $H_2O_2$ yield compared to pristine $C_3N_4$ [15]. P incorporation modified the band structure and induced interlayer spacing expansion, improving charge transport and reactant accessibility. It is also reported that the doping with O and P heteroatoms avoids the photodecomposition of $H_2O_2$, preventing the $4e^-$ oxidation from $O_2 \rightarrow H_2O_2 \rightarrow H_2O$ and echoing in a 11-fold increase in $H_2O_2$ yield [44]. Sulfur doping contributes to bandgap narrowing, enhancing visible light absorption, and increasing the active sites for $O_2$ activation, favoring the selectivity of the $2e^-$ ORR, reflected in the 16.4-fold increase in $H_2O_2$ yield [37]. This trend in $H_2O_2$ yield is further corroborated by the apparent quantum yield (AQY) values at 365 and 420 nm, which provide a quantitative measure of photocatalytic efficiency under different monochromatic excitation wavelengths (Figure 3c). The pristine $C_3N_4$ exhibited the lowest AQY values of 0.68 and 0.49% at 365 and 452 nm, respectively, reinforcing its inherent limitations in charge utilization. In contrast, heteroatom doping significantly enhanced the AQY across all cases. B- $C_3N_4$ exhibited AQY values of 5.05 and 4.74%, confirming the improved efficiency resulting from band structure modulation. O- $C_3N_4$ achieved AQY values of 8.12 and 7.30%, confirming its superior charge separation and light absorption properties. P- $C_3N_4$ exhibited AQY values of 5.86 and 5.43%, supporting the role of P doping in enhancing charge transport and minimizing $H_2O_2$ decomposition. Notably, S- $C_3N_4$ showed the highest AQY values of 8.55 and 8.08%, aligning with its superior performance in $H_2O_2$ generation due to enhanced visible-light absorption and optimized reaction selectivity. Overall, these results underscore the importance of heteroatom doping in tuning the photocatalytic performance of $C_3N_4$ and the selectivity of the reaction.

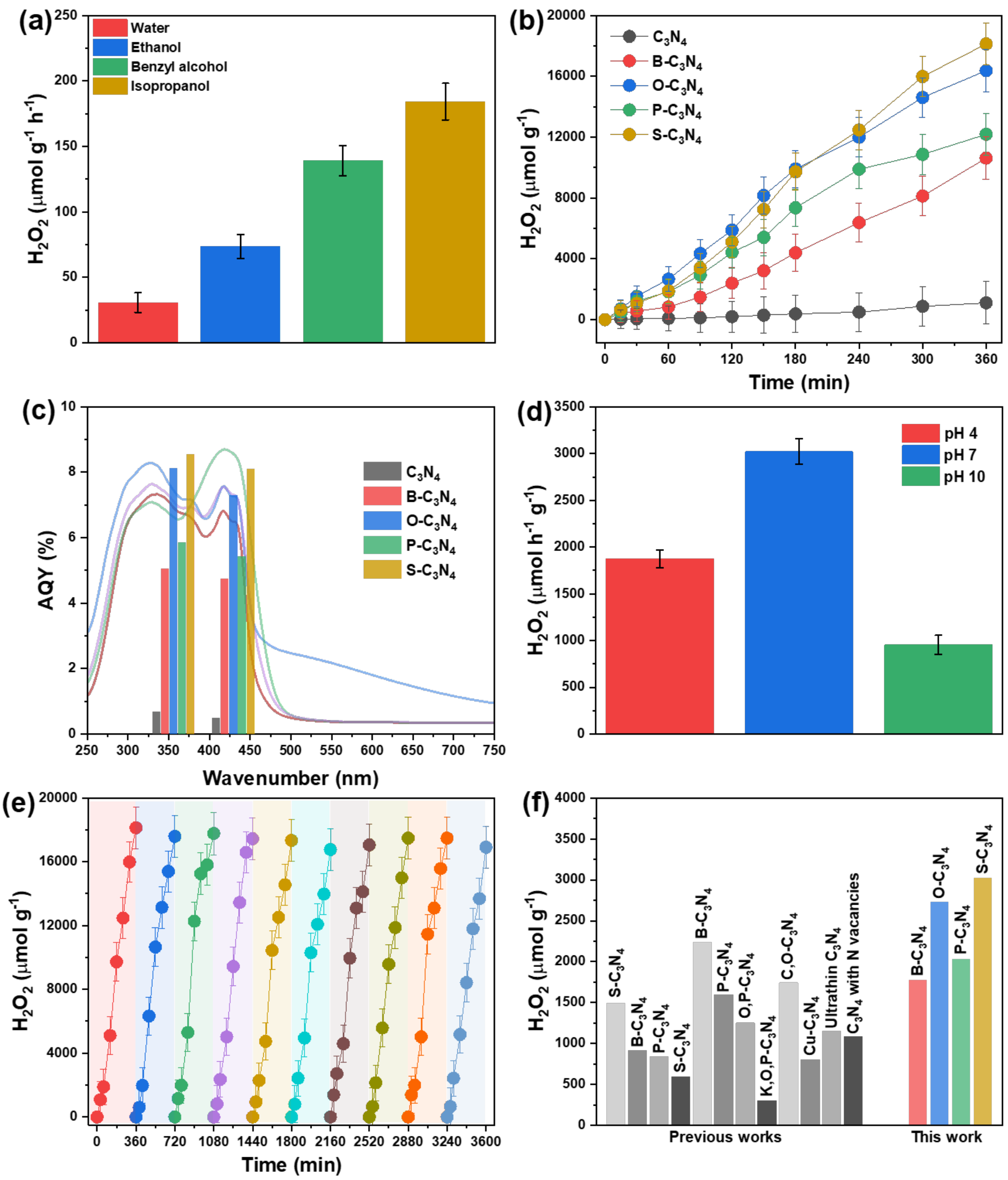

**Figure 3.** Effect of **(a)** sacrificial agent (SA) on $H_2O_2$ yield ([SA]: 10 wt. %, time: 6h; $C_3N_4$ dosage: 1 g $L^{-1}$); **(b)** $H_2O_2$ production kinetics for all the doped catalysts ([IPA]: 10 wt. %, time: 6h; pH 7; catalyst dosage: 1 g $L^{-1}$); **(c)** absorbance and AQY of $C_3N_4$, B-$C_3N_4$, O-$C_3N_4$, P-$C_3N_4$, and S-$C_3N_4$; **(d)** effect of pH on $H_2O_2$ yield ([IPA]: 10 wt. %, time: 6h; S-$C_3N_4$ dosage: 1 g $L^{-1}$); **(e)** reusability of S-$C_3N_4$ over 5 consecutive cycles ([IPA]: 10 wt. %, time: 6h; S-$C_3N_4$ dosage: 1 g $L^{-1}$); **(f)** $H_2O_2$ maximum yields reported for $C_3N_4$-based photocatalysts in comparison with the heteroatom-doped $C_3N_4$ catalysts studied in this work [9,39,48–55].

The $H_2O_2$ yield is significantly influenced by the pH, affecting both charge transfer dynamics and the reaction pathway (Figure 3d). At pH 7, the production yield is maximized (3022.1 μmol $g^{-1}$ $h^{-1}$), while at pH 4 and pH 10 the $H_2O_2$ yield is reduced to 1873.3 and 954.4 μmol $g^{-1}$ $h^{-1}$, respectively). The mechanism responsible for the optimal pH is related to the 1$e^-$ reduction of $O_2$ to generate superoxide ($O_2^{\bullet-}$) radicals, which is pH-dependent. At acidic pH, abundant protons favor $HO_2^{\bullet-}$ generation and accelerate side reactions. On the other hand, at alkaline pH, $O_2^{\bullet-}$ predominates, but $H_2O_2$ undergoes base-catalyzed disproportionation and radical-induced degradation into $H_2O$ and $O_2$ as it is less stable.

In addition, the ORR pathway shifts toward the 4e$^-$ reduction instead of to the desired 2e$^-$ pathway. At neutral pH, the best balance is achieved: the generation of $O_2^{\bullet-}$ is favored, as the p$K_a$ of $HO_2^{\bullet-}$ is 4.8, leading to the stabilization of intermediaries to the production of $H_2O_2$. Also, there is minimized decomposition due to the higher stability of $H_2O_2$ [56].

The stability of a photocatalyst over multiple cycles is a crucial factor in its practical applicability (Figure 3e). In the case of S-$C_3N_4$, the $H_2O_2$ production shows remarkable stability over ten consecutive cycles, with only a slight decline in yield from 3022.1 μmol g$^{-1}$ h$^{-1}$ in the first cycle to 2818.7 μmol g$^{-1}$ h$^{-1}$ in the tenth cycle. This minor yield loss of 6.7% demonstrates excellent durability and resistance to deactivation. This stability is related to: *(i)* structural and chemical stability of the catalyst, reinforced by the presence of sulfur – strengthening π-conjugation and improving charge carrier dynamics [57]. This is well demonstrated in the XRD pattern of S-$C_3N_4$ after ten consecutive cycles, presenting minimal differences regarding the pristine catalyst (Figure S16) – there is a decrease of intensity in the main peak at 27.8° and the disappearing of the less intense peaks at $2\theta < 15°$; *(ii)* effective charge separation, as discussed previously; *(iii)* resistance to self-$H_2O_2$ decomposition due to the excess of ROS [58]; and *(iv)* availability of active sites, as the catalyst does not accumulate inhibitory species on its surface.

The photocatalytic production of $H_2O_2$ using heteroatom-doped $C_3N_4$ has been widely studied, with various modifications enhancing efficiency by improving charge separation and surface reaction sites. In this work, B-, O-, P-, and S-doped $C_3N_4$ catalysts achieved yields of 1771.3, 2730.1, 2031.0, and 3022.1 μmol g$^{-1}$ h$^{-1}$, respectively, surpassing many previously reported values in the literature (Figure 3f and Table 2).

**Table 2.** Previous works reporting $C_3N_4$-based catalysts for the photocatalytic $H_2O_2$ production.

| Catalyst | Light source | Photon flux (photon s$^{-1}$ x10$^{19}$) | pH | $H_2O_2$ yield (μmol g$^{-1}$ h$^{-1}$) | Ref. |
|---|---|---|---|---|---|
| $C_3N_4$ with N vacancies | Solar light, 8 W | 4.7 | - | 1150 | [54] |
| Hierarchical $C_3N_4$ | Visible light, 300 W | 1.8 | 3 | 1083 | [55] |
| B-$C_3N_4$ | | 6.4 | 6.8 | 2240 | [49] |
| O,P-$C_3N_4$ | | 6.7 | 7 | 1250 | [9] |
| K,P,O-$C_3N_4$ | | 6.6 | 6.5 | 302 | [51] |
| C,O-$C_3N_4$ | | 1.8 | 7 | 1742 | [52] |
| P-$C_3N_4$ | | 1.8 | 7 | 1596 | [39] |
| S-$C_3N_4$ | | 1.9 | 6.8 | 1491 | [48] |
| Cu-$C_3N_4$ | | 1.7 | 7 | 800 | [53] |
| B-$C_3N_4$ | Visible light, 0.04 W cm$^{-2}$ | 1.8 | 7 | 1771 | This work |
| O-$C_3N_4$ | | | | 2730 | |
| P-$C_3N_4$ | | | | 2031 | |
| S-$C_3N_4$ | | | | 3022 | |

B-$C_3N_4$ prepared in the present work achieves similar and superior performance when compared to previous B-doped $C_3N_4$-based catalysts: 916 and 2240 μmol g$^{-1}$ h$^{-1}$ [49,50]. In the first case, the nanotube-like morphology of the catalyst can explain the different yield [49], while in the second case the high yield is related to the use of photoelectrocatalysis instead of single photocatalytic processes, suggesting a synergistic effect of these two techniques but increasing the energy demand of the process [50]. O-$C_3N_4$ and P-$C_3N_4$ synthesized in this work surpass the yields of O,P- and K,O,P-$C_3N_4$ catalysts from literature (1250 and 302 μmol $g^{-1}$ $h^{-1}$) [49,51]. Similar to the previous report, the nanotube-like morphology of the catalyst can explain the different yield [49], while in the other work, the results suggest that single O- and P-doping is more beneficial for enhancing efficiency [39,51]. S-$C_3N_4$ outperforms reported 1491 and 595 μmol $g^{-1}$ $h^{-1}$ obtained in previous studies [48,49]. The superior performance may be related to the sulfur introduction on the surface of $C_3N_4$ by electrostatic interactions [48] or by the nanotube-like morphology of the catalyst [49], while in the present work, nanosheets were obtained with sulfur replacing nitrogen atoms in the structure. Other reported catalysts, such as Cu-$C_3N_4$ (800 μmol $g^{-1}$ $h^{-1}$), ultrathin $C_3N_4$ (1150 μmol $g^{-1}$ $h^{-1}$), and $C_3N_4$ with N vacancies (1083 μmol $g^{-1}$ $h^{-1}$), also exhibit lower yields than the best-performing catalysts in this study. This highlights the superior efficiency of B-, O-, P-, and S-doping strategies, which likely enhance charge separation, increase active sites, and suppress side reactions that could lead to $H_2O_2$ decomposition. The notably higher performance achieved in this work is thus attributed to optimized doping strategies that introduce mid-gap states, modify band structures, and improve charge carrier dynamics.

### 2.4. Charge transfer dynamics and mechanism of enhanced performance

The charge transfer dynamics of the pristine $C_3N_4$ and the heteroatom-doped catalysts were investigated by scavenger studies, steady-state photoluminescence spectroscopy (PL), photocurrent measurements (PC), and electrochemical impedance spectroscopy (EIS) (Figure 4).

The presence of scavengers influences the $H_2O_2$ production, as they selectively quench ROS involved in the reactions. The differences in the $H_2O_2$ yield provide insights into the dominant reaction pathways and charge carrier dynamics (Figure 4a). Without scavengers, $H_2O_2$ yield would be limited by charge recombination and competing side reactions, leading to a yield of 778.7 μmol $g^{-1}$ $h^{-1}$. The minor roles are attributed to hydroxyl ($^{\bullet}OH$) and superoxide ($O_2^{\bullet-}$) radicals. Quenching $^{\bullet}OH$ and $O_2^{\bullet-}$ significantly increase $H_2O_2$ yield (3591.6 and 2970.1 μmol $g^{-1}$ $h^{-1}$, respectively), confirming that this radical primarily acts as a recombination center rather than active intermediate, promoting side reactions rather than the selective $2e^-$ ORR pathway [59]. An intermediate role is assigned to electrons ($e^-$), resulting in a $H_2O_2$ yield of 1019.8 μmol $g^{-1}$ $h^{-1}$. A moderate yield was noted for $e^-$ quenching, as well as a slight increase compared with the absence of scavengers, confirming that electrons are necessary but not a key factor in the $H_2O_2$ production. Although electrons are necessary for $O_2$ reduction, they can reduce $O_2$ to $O_2^{\bullet-}$, which leads to undesired side reactions [60]. On the opposite, major roles are attributed to singlet oxygen radical ($^1O_2$) and holes ($h^+$). By quenching these active species, $H_2O_2$ yields of 427.0 and 3022.1 μmol $g^{-1}$ $h^{-1}$ are achieved, respectively. When $^1O_2$ is quenched, a drastic reduction on the yield is observed, confirming its major role as key intermediate in the $2e^-$ ORR pathway [61]. The presence of excessive $h^+$ promotes $e^-/h^+$ recombination, reducing the efficiency. Holes can also participate in the WOR forming $^{\bullet}OH$, which does not contribute to $H_2O_2$ production and may instead degrade it [62]. When $h^+$ is quenched, an increase in the yield is observed, showing that removing these active species improves charge separation and the photocatalytic efficiency. The formation of $^1O_2$ and $^{\bullet}OH$ was estimated by quantifying the concentration of N, N-p-nitrosodimethylaniline and 2-hydroxyterephthalic acid over time, respectively (Figure S17). All the materials show the same trend, as the concentration of $^1O_2$ being significantly higher than that of $^{\bullet}OH$. When the pristine catalyst was modified with heteroatoms, the amount of ROS increased in all cases, demonstrating the ability of the modified materials to generate ROS. As expected, the ROS generation follows the same tendency of $H_2O_2$ yields, which agrees with the PL, EIS, and photocurrent results discussed in this section.

Photoluminescence (PL) provides crucial insights into the charge carrier dynamics and recombination behavior of $C_3N_4$ and its heteroatom-doped derivatives (Figure 4b). The emission spectra were deconvoluted using Gaussian fitting, with four contributions observed across all samples. The quality of the fitting was confirmed by a $R^2$ higher than 0.99 for all cases. The four contributions are: (i) a high-energy emission at 3.35 eV (UV emission), ascribed to direct $e^-/h^+$ recombination, due to near band edge emission; (ii) a strong peak centered at 2.90 eV (violet), associated to $\pi \rightarrow \pi$ transitions within the conjugated heptazine units; (iii) a band at 2.75 eV (blue), linked to $n \rightarrow \pi^*$ transitions involving lone-pair electrons of nitrogen atoms in the π-conjugated units [63]; (iv) a lower-energy emission at 2.35 eV (blue-to-green), related to defect-related states such as N vacancies and structural distortions [64]. Pristine $C_3N_4$ exhibits a strong PL emission, confirming the fast $e^-/h^+$ recombination (Figure 4c). B- (Figure 4d) and P-doping (Figure 4f) introduce mid-gap states that slightly redshift the $\pi \rightarrow \pi$ (from 2.95 eV in pristine $C_3N_4$ to 2.97 in B- and P-doped) and $n \rightarrow \pi^*$ transitions (from 2.73 to 2.77 eV) [65]. By being less electronegative than nitrogen and carbon, B and P introduce acceptor-like states that distort the π-conjugated heptazine framework and shift the $\pi \rightarrow \pi$ and $n \rightarrow \pi^*$ transitions as carriers can recombine via these mid-gap states [66]. On the other side, O- (Figure 4e) and S-doping (Figure 4g) enhance delocalization of photogenerated carriers, reducing recombination and resulting in pronounced PL quenching. By replacing C in the heptazine framework, they increase the density of states near the CB and provide new transport pathways, reducing recombination [14]. The shifts in the relative contributions of the four deconvoluted bands are linked to the increased structural disorder and interlayer spacing [67], supported by XRD and FTIR data. The lower emission intensity observed for O- and S-doped $C_3N_4$ suggests improved charge separation, which correlates with their higher photocatalytic performance.

The photocurrent response provides insight into the charge separation efficiency of the photocatalysts (Figure 4h). Pristine $C_3N_4$ exhibits a moderate photocurrent, indicating a limited ability for the photogenerated charge carrier's separation, mainly due to rapid $e^-/h^+$ recombination [68]. Upon B-, O-, P-, and S-doping, the photocurrent response increases significantly, reflecting enhanced charge carrier generation and reduced recombination. S-doped $C_3N_4$ exhibits the highest photocurrent response among the doped materials, demonstrating improved conductivity and more efficient charge transport [9]. The enhanced photocurrent directly correlates with improved photocatalytic activity, as more charge carriers are available for the reaction.

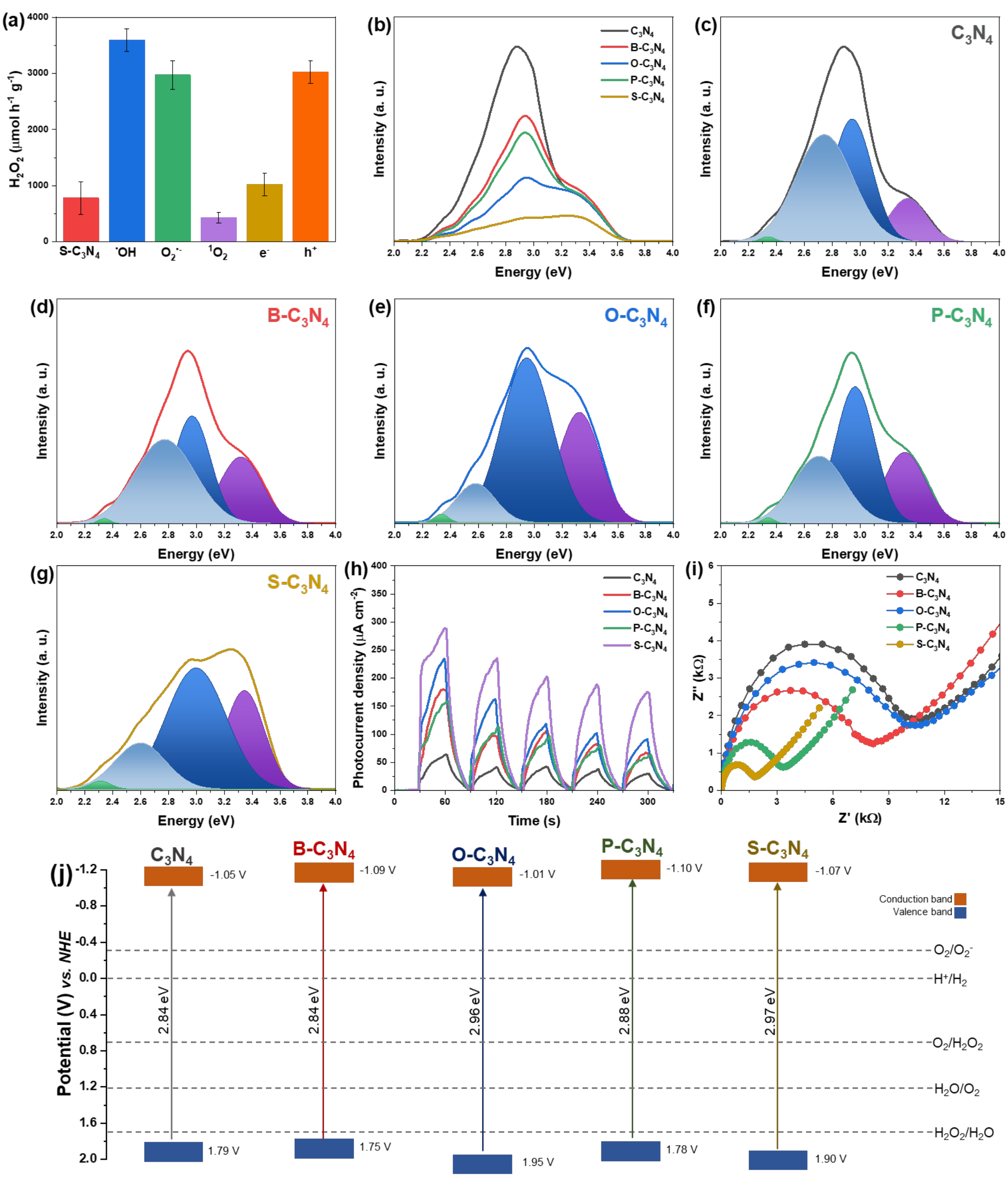

**Figure 4.** **(a)** Effect of different scavengers on $H_2O_2$ yield ([scavenger]: 10 wt. %; time: 6h; $S\text{-}C_3N_4$ dosage: 1 g $L^{-1}$); **(b)** PL emission spectra of all catalysts; deconvoluted PL emission spectra of **(c)** $C_3N_4$; **(d)** $B\text{-}C_3N_4$; **(e)** $O\text{-}C_3N_4$; **(f)** $P\text{-}C_3N_4$; and **(g)** $S\text{-}C_3N_4$; **(h)** photocurrent measurements and **(i)** EIS Nyquist plots for $C_3N_4$, $B\text{-}C_3N_4$, $O\text{-}C_3N_4$, $P\text{-}C_3N_4$, and $S\text{-}C_3N_4$; **(j)** schematic band edge positions of catalysts alongside the redox potentials.

The electrochemical behavior, as analyzed by the Nyquist plots, further supports these findings by revealing a reduction in charge transfer resistance for the doped materials (Figure 4i). The diameter of the semicircle in the Nyquist plots is notably reduced upon doping compared with the pristine catalyst, indicating faster charge transfer kinetics [69]. The trend follows $S\text{-}C_3N_4 < O\text{-}C_3N_4 < P\text{-}C_3N_4 < B\text{-}C_3N_4 < C_3N_4$, where S- and O-doping lead to the most significant reduction in resistance. The lower resistance confirms improved $e^-$ mobility and interfacial charge transfer, which are essential for enhancing photocatalytic efficiency [68].

The flat band was estimated by Mott-Schottky plots (Figure S18). The positive slope determines the n-type semiconductor type of $C_3N_4$-based catalysts [54]. As the flat band for n-type semiconductor lies 0.1 eV below the conduction band (CB), the CB positions of $C_3N_4$, B-$C_3N_4$, O-$C_3N_4$, P-$C_3N_4$, and S-$C_3N_4$ were established as -1.15, -1.19, -1.11, -1.20, and -1.17 V, respectively. Combining these data with the band gap energy, the valence band was calculated for all the catalysts. The band positions *vs.* RHE (pH = 0) are provided in Figure 4(j). The band alignment for all the catalysts is aligned with the redox potential of the reactions that generate ROS ($O_2/O_2^-$, $H^+/H_2$, $O_2/H_2O_2$, $H_2O/O_2$, and $H_2O_2/H_2O$). Despite not including the redox potential of $^{\bullet}OH$ generation (2.72 V), this radical can be generated through side reactions (water oxidation, hydroxide oxidation, hydrogen peroxide oxidation, and superoxide disproportionation) [70].

## 3. CONCLUSIONS

In this work, $C_3N_4$ was modified with B, O, P, and S heteroatoms enhancing its structural, electronic, and optical properties, and improving its performance for photocatalytic $H_2O_2$ production.

A few-layered structure with stacked nanosheets was achieved for pristine and doped $C_3N_4$. The doping resulted in a reduction of crystallinity and an increase of interlayer spacing, and shifts in binding energies indicated modified electronic environments. Suitable optical properties for photocatalysis under sunlight were observed for all the catalysts, even considering the increased band gap energy for O- and S-doped catalysts (from 2.84 to 2.96 and 2.97 eV, respectively). The impact of doping $C_3N_4$ with B, P, O, and S heteroatoms depended strongly on both the heteroatom and its substitution site. B substitutes preferentially at C sites inducing local structural flattening and separating spatially HOMO and LUMO states into different layers. P enters as PH at C sites, preserving structure but causing HOMO and LUMO wavefunctions to shrink around the dopant. O prefers C substitution and, despite inducing in-plane deformation, preserves a delocalized LUMO, while modifies the HOMO wavefunction behavior. S atoms prefer substituting C, bending the nanosheets in-plane and enhancing delocalization of both HOMO and LUMO states.

Photocatalytic $H_2O_2$ production demonstrated that heteroatom doping significantly enhanced efficiency: pristine $C_3N_4$ achieved a yield of 184.4 µmol $g^{-1}$ $h^{-1}$, while B-, O-, P-, and S-doped $C_3N_4$ achieved 1771.3, 2730.1, 2031.0, and 3022.1 μmol $g^{-1}$ $h^{-1}$, respectively, as a result of improved charge separation, modulated band structures, and the presence of active sites favoring the $2e^-$ ORR pathway. The same trend was observed for the AQY, with an increase from 0.68% for pristine $C_3N_4$ to 5.05, 8.12, 5.86, and 8.55% for B-, O-, P-, and S-$C_3N_4$ catalysts. The effect of pH showed a maximized yield under neutral conditions, while extreme pH values promoted decomposition or unfavorable reaction pathways. The S-doped catalyst demonstrated excellent stability over ten cycles, with minimal loss in performance (< 10%). Charge carrier dynamics studies confirmed reduced recombination rates and improved charge separation for doped catalysts. Scavenger tests identified $^1O_2$ and $h^+$ as key contributors to the reaction.

The combination of structural modifications and optimized charge carrier dynamics underscores the potential of these catalysts for sustainable $H_2O_2$ production. Overall, this study provides valuable insights into heteroatom doping to reach record-breaking performances, paving the way for further advancements in photocatalysis, sustainable solar-to-chemical conversion, and green energy.

## 4. EXPERIMENTAL

The experimental details are described in the Section 4 of Supporting Information.

### Acknowledgements

This work received support from the Horizon Europe Framework Programme (HORIZON-TMA-MSCA-SE), project No. 101131229, “Piezoelectricity in 2D-materials: materials, modeling, and applications (Piezo2D). The authors thank the financial support from the Spanish Agencia Estatal de Investigación (AEI) through National Fund for Scientific and Technological Development (FONDECYT REGULAR-ANID, 1220088) and Tailing43Green-ERAMIN projects. This study forms part of the Advanced Materials program and was supported by MCIN with funding from European Union NextGenerationEU (PRTR-C17.I1) and the IKUR strategy of the Basque Government Education Department. Basque Government Industry and Education Departments under the ELKARTEK and PIBA (PIBA-2022-1-0032) programs are also acknowledged. H. Salazar thank the Ministerio de Economía y Competitividad for the Juan de la Cierva (JDC2022-050319-I) research contract. The authors thank the technical and human support provided by SGIker (UPV/EHU). I. Infante and J. Llusar acknowledge Horizon Europe EIC Pathfinder program through project 101098649-UNICORN. DFT calculations were carried at the Donostia International Physics (DIPC) Supercomputing Center, for which the authors acknowledge for the technical and human support. We also thank PRACE for awarding us access to Leonardo at CINECA, Italy.